\documentclass[journal,twoside,web]{ieeecolor}
\usepackage{generic}
\usepackage{cite}
\usepackage{amsmath,amssymb,amsfonts}
\usepackage{graphicx}
\usepackage{textcomp}
\usepackage{multirow}
\usepackage{amsmath}
\usepackage{stfloats}
\usepackage{color}

\usepackage{mathrsfs}

\usepackage{easyReview}
\usepackage[switch]{lineno}    
\usepackage[ruled]{algorithm2e}
\usepackage{verbatim}
\usepackage{booktabs}
\usepackage{epstopdf}

\def\BibTeX{{\rm B\kern-.05em{\sc i\kern-.025em b}\kern-.08em
    T\kern-.1667em\lower.7ex\hbox{E}\kern-.125emX}}
\begin{document}
\title{GlanceSeg: Real-time microangioma lesion segmentation with gaze map-guided foundation model for early detection of diabetic retinopathy}

\author{Hongyang Jiang\textsuperscript{1}, Mengdi Gao\textsuperscript{1}, Zirong Liu, Chen Tang, Xiaoqing Zhang, Shuai Jiang, Wu Yuan\textsuperscript{\dag}, and Jiang Liu\textsuperscript{\dag}  
\thanks{This work was supported in part by General Program of National Natural Science Foundation of China under Grant 82272086, Shenzhen Science and Technology Program under Grant KQTD20180412181221912 and Grant JCYJ20200109140603831, the Innovation and Technology Fund (ITF) of Hong Kong SAR (ITS/240/21), and the Science, Technology, and Innovation Commission (STIC) of Shenzhen Municipality (SGDX20220530111005039).}
\thanks{Hongyang Jiang, Xiaoqing Zhang, and Jiang Liu are with the Department of Computer Science and Engineering, Southern University of Science and Technology, Shenzhen, China. (e-mail: jianghy3@sustech.edu.cn; 11930927@mail.sustech.edu.cn; liuj@sustech.edu.cn).}
\thanks{Zirong Liu, Chen Tang, and Jiang Liu are with the School of Ophthalmology and Optometry and Eye Hospital, Wenzhou Medical University, Wenzhou 325027, China. (e-mail: zirongliu98@qq.com; tangchen419@163.com; liuj@sustech.edu.cn).}
\thanks{Shuai Jiang is with Intelligent Vision Plus Technology Co., Ltd., Shenzhen, China. (e-mail: jsiacb@foxmail.com).}
\thanks{Mengdi Gao and Wu Yuan are with the Department of Biomedical Engineering, The Chinese University of Hong Kong, Hong Kong Special Administrative Region of China. (e-mail: mengdigao@cuhk.edu.hk; wyuan@cuhk.edu.hk).}
\thanks{\textsuperscript{1}These authors contributed equally to this work.}
\thanks{\textsuperscript{\dag}These authors are co-corresponding authors.}
}

\maketitle

\begin{abstract}
Early-stage diabetic retinopathy (DR) presents challenges in clinical diagnosis due to inconspicuous and minute microangioma lesions, resulting in limited research in this area. Additionally, the potential of emerging foundation models, such as the segment anything model (SAM), in medical scenarios remains rarely explored. In this work, we propose a human-in-the-loop, label-free early DR diagnosis framework called GlanceSeg, based on SAM. GlanceSeg enables real-time segmentation of microangioma lesions as ophthalmologists review fundus images. Our human-in-the-loop framework integrates the ophthalmologist's gaze map, allowing for rough localization of minute lesions in fundus images. Subsequently, a saliency map is generated based on the located region of interest, which provides prompt points to assist the foundation model in efficiently segmenting microangioma lesions. Finally, a domain knowledge filter refines the segmentation of minute lesions. We conducted experiments on two newly-built public datasets, i.e., IDRiD and Retinal-Lesions, and validated the feasibility and superiority of GlanceSeg through visualized illustrations and quantitative measures. Additionally, we demonstrated that GlanceSeg improves annotation efficiency for clinicians and enhances segmentation performance through fine-tuning using annotations. This study highlights the potential of GlanceSeg-based annotations for self-model optimization, leading to enduring performance advancements through continual learning.

\end{abstract}


\section{Introduction}
\label{sec:introduction}

\IEEEPARstart{R}{etinal} pathologies contribute to millions of cases of blindness worldwide, with diabetic retinopathy (DR) being a major cause of vision impairment \cite{abramoff2010retinal}. In response, computer-aided diagnostic (CAD) systems for DR have emerged \cite{quellec2017deep,gulshan2016development,wei2023caudr}, leveraging extensive and meticulously annotated data. However, acquiring such data is costly and time-consuming.
Despite the impressive diagnostic performance of CAD systems for DR, their reasoning processes often lack transparency and may produce unexpected errors, leaving room for optimization in clinical applications. Furthermore, most CAD systems do not prioritize the challenging task of early detection of DR, which involves identifying small lesions like microangiomas. Therefore, there is a need to develop a clinically beneficial CAD system for early DR detection that reduces the reliance on annotations or even adopts a label-free approach.

In a clinical setting, an ophthalmologist's diagnostic process is essentially a meticulous collection of evidence, a practice often referred to as evidence-based medicine \cite{sackett1996evidence}. Specifically, when evaluating medical images, clinicians dedicate themselves to identifying any lesions or abnormalities present in the image to aid their decision-making process. The areas of the image that receive the most attention from clinicians are of paramount importance and have a direct correlation with the diagnostic outcome. Generally, clinicians can swiftly diagnose conditions when the images display evident lesions. However, the identification of suspected abnormalities can be time-consuming in instances where the lesions are subtle. Occasionally, hidden and small-scale lesions are likely to be missed, presenting a challenge in the current clinical landscape. This indicates that the integration of gaze maps, derived from the clinician's image evaluation process, as an attention mechanism could enhance the detection of lesion areas, particularly the small-scale lesions. Gaze maps provide clinicians' prior knowledge through a human-in-the-loop mechanism without imposing additional annotation workload on them, thereby holding the potential to decrease the rate of missed diagnoses.

Nowadays, the advent of foundation models such as the Segment Anything Model (SAM) \cite{kirillov2023segment} has prompted the medical imaging community to investigate methods to harness the exceptional generalization capabilities of these models to enhance the performance and reduce the training costs of specialized models \cite{qiu2023visionfm}. Existing studies \cite{huang2023segment,qiu2023large,shi2023generalist} underscore the remarkable potential of SAM for use in medical contexts. However, the direct application of SAM in the medical field often results in inconsistent segmentation results. Moreover, the influence of prompt points on the segmentation efficacy of SAM-based foundation models is substantial. Therefore, when integrating foundation models into medical applications, it is imperative to conduct specific investigations into effective prompt points to facilitate efficient target segmentation.

In this work, we propose an innovative approach that takes into account a clinician's diagnostic behavior by collecting the real-time eye-tracking data from clinicians during their diagnostic process, facilitated by an eye-tracker. This eye-tracking information serves as a bottom-up attention mechanism (i.e., medical prior knowledge) that mirrors the clinician's process of identifying medical evidence, and it can be easily digitized for computational purposes. Moreover, the process of generating eye-tracking data is unobtrusive and user-friendly for clinicians, assisting in the identification of regions of interest (ROI). Consequently, we incorporated this bottom-up attention mechanism into the development of the CAD system.
In addition, we utilized the saliency map as a bottom-up attention computation method. To exploit the zero-shot segmentation capability of the widely-used SAM, we introduced a prompt points generation method that employs an integrated saliency computation approach on the ROI. Furthermore, we integrated a domain knowledge filter to single out target lesions, which not only aids clinicians' diagnoses but also serves as weakly annotated masks for Artificial Intelligence models to learn from.

In summary, the contributions of this paper are as follows:
\begin{enumerate}[]
	\item This study presents a real-time, unsupervised medical image segmentation framework, specifically designed for small lesions, which utilizes a foundational model built on a multi-head attention mechanism.
	\item The gaze map was initially employed to support top-down attention, allowing for the approximate localization of areas of interest related to small lesions, as indicated by the clinician's eye movement..
	\item Subsequently, integrated saliency maps were derived to generate rational prompt points for the foundational model (e.g., SAM), thus forming the bottom-up attention mechanism.
	\item After that, a domain knowledge filter was utilized to post-process the lesion segmentation results, introducing an optimization technique based on prior medical domain knowledge.
	\item GlanceSeg facilities both annotation and diagnosis. Its effectiveness was validated using two restructured public datasets, specifically for the segmentation of microangiomas in early diabetic retinopathy detection.
\end{enumerate}	

\begin{figure*}[t!]
	\centering
	\includegraphics[width=1.8\columnwidth]{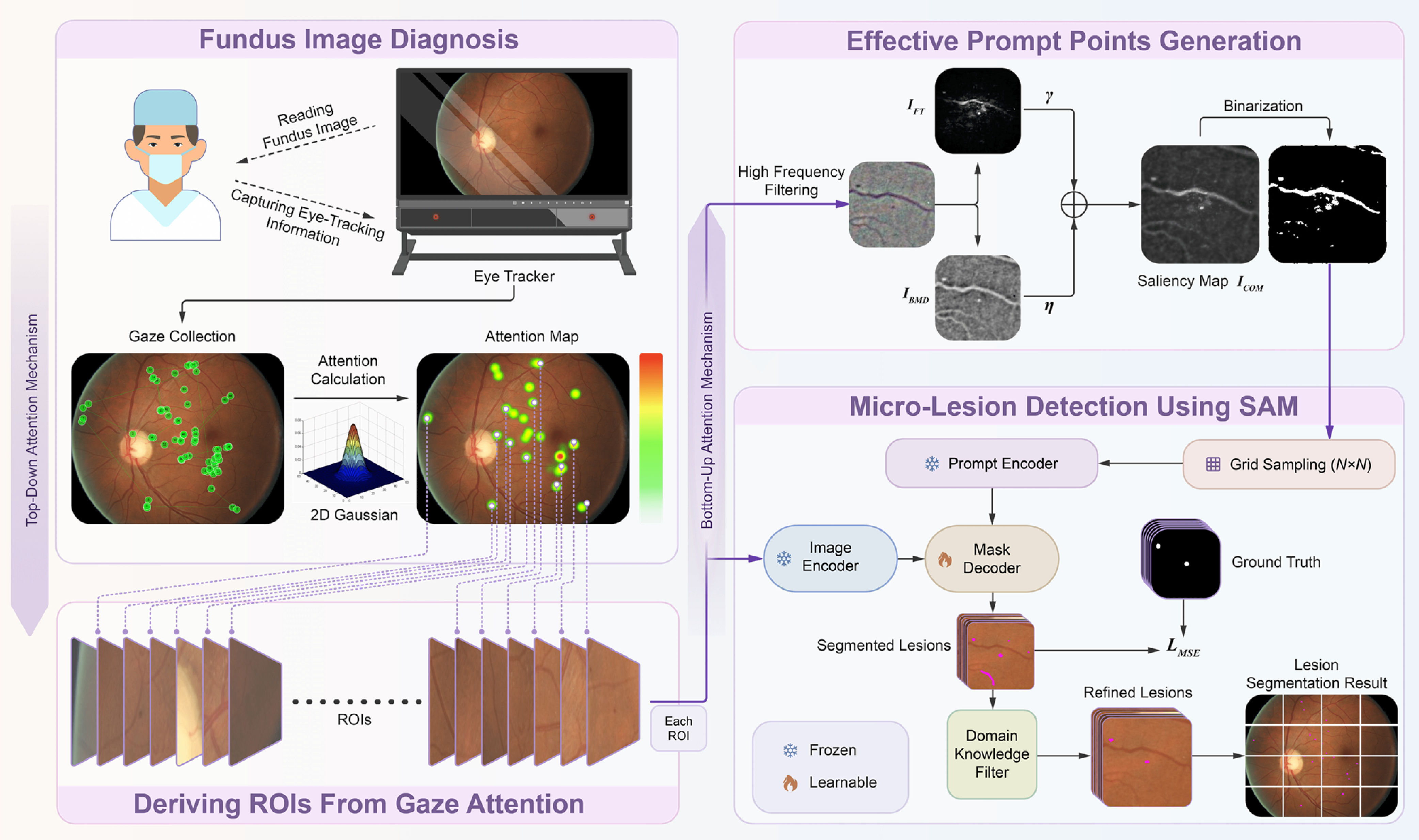}
	\vspace{0em}
	\caption{Pipeline of the proposed real-time gaze map-guided early DR detection framework based on SAM (i.e., GlanceSeg). It integrates gaze-map as top-down attention and saliency map as bottom-up attention, enabling real-time segmentation of small lesions while clinicians review the images. Optionally, the GlanceSeg's performance can be enhanced through fine-tuning with the annotations. }
	\label{fig:method}
	\vspace{-1.4em}
\end{figure*}

\section{Related work}
\subsection{Gaze map-guided medical image analysis}
In accordance with the principles of evidence-based medicine \cite{sackett1996evidence}, the diagnostic process entails seeking corroborative findings from imaging to inform decision-making for individual patients. Abnormalities in medical images that are pertinent to decisions are often subtle and occupy a small number of pixels, especially when compared to natural images. This challenge may be further exacerbated by the limited size of medical image data. Consequently, researchers are exploring the use of clinicians' gaze maps, generated during the reading and diagnosis of medical images, as a strategy to mitigate these difficulties. For example, Hong et al. \cite{jiang2023eye} utilized the weighted gaze map as supervised mask to guide the learning of attention based DNNs. Their experimental results validated that gaze maps could be considered as medical prior knowledge to improve both the accuracy and interpretability of early DR detection model based on lightweight training images. Moreover, Wang et al. \cite{wang2022follow} introduced a gaze-guided attention model to grade the osteoarthritis into four categories based on X-ray images, which obtained better interpretability and classification performance. Based on radiologists’ gaze maps, Karargyris et al. \cite{karargyris2021creation} developed a U-Net-based multi-head model to concurrently predict experts’ attention and classification results in distinguishing congestive heart failure, pneumonia, and normal cases. 

On the other hand, eye-gaze maps can be effectively utilized in the field of automated medical image annotation. Stember et al. \cite{stember2019eye} presented an eye-tracking-based dynamic annotation approach for identifying lesions and organs in multi-modal medical images, including computed tomography (CT) and magnetic resonance and imaging (MRI). Furthermore, Stember et al. \cite{stember2020integrating} combined eye-tracking and speech information together to annotate abnormal regions in medical images. Wang et al. \cite{wang2023gazesam} innovatively presented the GazeSAM system with the eye-gaze data as the input prompt for SAM, which automatically generated the segmentation mask in real time. Although gaze maps can act as indicators of visual attention, their direct use as prompts may be too coarse and imprecise. In this study, we utilized gaze maps acquired from clinicians during their review process to create a user-friendly, human-in-the-loop CAD system for early-stage DR. These gaze maps function as a top-down attention mechanism, aiding in the rough localization of regions of interest that contain small and concealed microangiomas.

\subsection{Segment anything model applied in the medical field}
SAM \cite{kirillov2023segment} is a foundation model that has been extensively trained on more than one billion annotations. Its training primarily focuses on natural images, and it excels in interactively segmenting user-defined objects of interest across different prompt settings. However, when it comes to medical imaging, SAM faces distinct challenges due to the specific nature of this domain. Numerous researchers are dedicated to investigating the application of SAM in medical fields. Mazurowski et al. \cite{mazurowski2023segment} conducted a thorough assessment of SAM’s capacity for medical image segmentation on a collection of 19 medical imaging datasets from different modalities and anatomies. They discovered that SAM’s performance based on single prompts highly varies depending on the dataset and the task, and SAM demonstrates enhanced performance when utilizing box prompts as opposed to point prompts. Besides, they reported that SAM shows impressive zero-shot segmentation performance for specific medical imaging datasets, but moderate to suboptimal performance for others. Huang et al. \cite{huang2023segment} constructed a comprehensive medical segmentation dataset comprising 16 modalities, 68 objects, and 553K slices that cover the entire body, and investigated various testing strategies for SAM. The authors demonstrated that SAM’s zero-shot segmentation capability is inadequate for immediate application in the medical field. Moreover, they observed considerable variation in SAM’s performance across different modalities and objects, even for distinct modalities of the same object, underscoring its instability. Additional studies \cite{he2023accuracy, mattjie2023exploring, deng2023segment} have also evaluated SAM’s performance in the medical field. The consensus among these investigations is that while SAM holds promise as a universal medical image segmentation model, its current performance remains inconsistent. Therefore, researchers should prioritize the task of fine-tuning SAM to improve its reliability and ensure more consistent results when applied to medical images.

To explore the potential of SAM within the unique context of medical imaging, researchers have initiated the process of transferring the generalized knowledge from the foundation model into specialized medical models. Ma et al. \cite{ma2023segment} adapted SAM for general medical image segmentation through fine-tuning the decoder of SAM, and the resulting performance of MedSAM exceeded that of the default SAM.
PromptUNet \cite{wu2023promptunet} developed a novel prompt-based medical segmentation model through expanding the existing prompts in SAM to supportive and En-face prompts. The authors have demonstrated that PromptUNet surpassed a range of state-of-the-art medical image segmentation methods, as well as those tailored SAM models for the medical field \cite{wu2023medical, ma2023segment}. Qiu et al. \cite{qiu2023learnable} proposed the learnable ophthalmology SAM, suitable for segmentation of multiple targets in multimodal ophthalmic images. Specifically, they incorporated a learnable prompt layer between each transformer layer and substituted the mask decoder with a task-specific segmentation head. During the training phase, only the prompt layer and task-specific head were trained while the transformer layers and patch embedding were frozen. This approach demonstrated superior performance in segmenting blood vessels, lesions, and retinal layers compared to the default SAM. Wang et al. \cite{wangsammed} leveraged the SAM to tackle the medical image annotation task by automatically generating the input prompt through a trainable prompt generation module.

The current research progress represents merely an initial step in the application of foundation models in healthcare. When transferring a generalist foundation model to a medical-specific application, it is essential to consider the characteristics of medical images, such as the unique anatomical structures and abnormalities, the image resolution, and the relatively small lesions. Moreover, the strategic selection of appropriate prompt points plays a crucial role in achieving more precise segmentation of the target lesions. In this work, we employed a saliency map derived from gaze map-guided segmented ROI. This approach allowed us to generate adaptive prompt points, thereby facilitating the segmentation of microangiomas in the early stages DR.

\section{MATERIAL AND METHODS}
In this work, we propose GlanceSeg, an eye-tracking-guided micro lesion real-time segmentation method built on the fundation model (SAM). First, Section A introduces the eye-tracker-based clinical diagnostic environment and related configurations. Second, Section B delves into the details of multiple attention calculation approaches, including top-down attention and bottom-up attention. Next, Section C 
elaborates on the process of prompt generation and lesion segmentation using the SAM model. Finally, Section D explains the the domain knowledge filtering module, which plays a crucial role in optimizing lesion segmentation performance.

\subsection{Clinical diagnosis setting \& eye-tracking acquisition}\label{subsec9}
The desktop eye-tracker, i.e., Tobii Pro Spectrum, was essential equipment for medical image reviewing and diagnosis in our clinical settings. The Tobii Pro Spectrum with two eye-tracking cameras was located below the LED monitor, and provided optimal precision with an root mean square (RMS) of about $0.01^{\circ}$ and a sampling frequency of up to 1200Hz. Based on the physiological response of human eyes \cite{andersson2010sampling}, a sampling of 60Hz can meet the requirements. Before reviewing the image, the clinician needs to sit within the appropriate distance range from the eye-tracker, i.e., 55 to 70 cm. To reduce the eye-tracking error, every clinicians performed a 9-point calibration procedure at their first use. During the image reviewing, a series of gaze points on each diagnostic image were recorded in real time.

\subsection{Construction of multiple attention mechanism}\label{subsec10}
\subsubsection{Top-down attention computation}\label{subsec11}
Following the collection of gaze points, we implemented a simulated attention calculation approach. However, it is important to note that even after personalized calibration, slight deviations in the clinician's gaze points may occur due to the inherent system error of the instrument. We first introduced a Gaussian function $G(x,y)$ (shown as Eq.\ref{eq1}) to transform each gaze point into the gaze area.
\begin{equation}
	G(x, y)=\frac{1}{\sqrt{2 \pi} \sigma} e^{\frac{-\left[\left(x-x_{c}\right)^{2}+\left(y-y_{c}\right)^{2}\right]}{2 \sigma^{2}}}
	\label{eq1}
\end{equation}
$(x_c, y_c)$ represents the central gaze point, and the pixel-level distance of a variance $\sigma$ is considered as the effective range of the field of view. 
We can acquire a proportional equation as follows: 
\begin{equation}
	\frac{\pi (\sigma^{'})^{2}}{H\cdot  W} = \frac{\pi \sigma^{2}}{H_p\cdot  W_p}
	\label{eq2}
\end{equation}
where $H_p=1080$ and $W_p=1920$ represent the resolution of the monitor, and $H=48$ and $W=64$ are the physical values corresponding to $H_p$ and $W_p$, respectively. Based on the geometric operation, $\sigma^{'} =\theta / 360^{\circ}$, where $\theta$ is regarded as visual angle error. Then, the pixel variance $\sigma_p$ on the monitor can be estimated as follows:
\begin{equation}
	\sigma =\frac{\theta }{360^{\circ }}\pi R\sqrt{\frac{H_p\cdot  W_p}{H\cdot  W} } 
	\label{eq3}
\end{equation}
where $R$ is the approximate distance between the eyeball and the fixation point on the screen. In our experiments, $\theta$ is set to $1^{\circ}$, which can offset the majority of errors. Finally, the calculated value of $\sigma$ is in 18.69 (R=55cm) $\sim$ 23.79 (R=70cm) and thus we pre-set $\sigma$ to 25 for better overcoming the errors. Then, we can obtain the corresponding top-down attention map (i.e., gaze map) of the diagnostic image.

According to the gaze map, we initially proposed an attention dispersion score (ADS) to measure the focusing degree of clinicians' attention. The computation of ADS aims to to minimize the following formula:
\begin{equation}
	\left\{
	\begin{array}{llll}
		f^*_{ADS}(x^*,y^*) = \min\limits_{(x^*,y^*)} \frac{ \sum\limits_{i=1}w_i\left\| (x_i,y_i) - (x,y) \right\|_2}{Z} \\
		Z = \sum\limits_{i=1}w_i \cdot \frac{\sqrt{H\cdot W}}{100} \\
		0 \leq x \leq H \\
		0 \leq y \leq W \\
	\end{array}
	\right.
\end{equation}
where $w_i$ denotes the attention value of the valid gaze point $(x_i,y_i)$ on the gaze map, $(x^*,y^*)$ represents the clustering center point of all valid gaze points, and $Z$ is the normalization factor. Inside, the ADS $f^*_{ADS}$ of each gaze map can be calculated accordingly.

\subsubsection{ROI extraction \& enhancement with bottom-up attention}\label{subsec12}
Clinically, doctors usually carefully observe important areas and suspicious signs of lesions for a long time. Thus, in order to extract the ROI, we implemented a series of image processing operations. (1) Binary the whole top-down attention map to determine the range of ROI. (2) Calculate the center point of each ROI. (3) Crop each sub-image of $M\times M$ size based on each ROI center point. Notably, the length of M should be proportional to the ROI area so as to ensure that the ROI is completely in the sub-image.

Generally, the styles of different local regions from different images can be quite different, which increases the difficulty of algorithm adaptation. Considering this issue, we conducted an popular image enhancement approach based on an Gaussian filter template. The enhanced ROI image $I_{Enh}$ can be computed as:
\begin{equation}
	I_{Enh}=\alpha \cdot I_{ROI} + \beta \cdot [I_{ROI}\otimes G_{FT}] + \lambda
	\label{eq4}
\end{equation}
where $I_{ROI}$ and $G_{FT}$ represents original cropped ROI image and Gaussian filter template, respectively. In addition, $\alpha$,  $\beta$ and $\lambda$ are weighted parameters that control the saturation and brightness of the image. Referring to the relevant literature \cite{jiang2019interpretable}, the empirical values of $\alpha$,  $\beta$, and $\lambda$ are 4, -4, and 128.

\subsubsection{Bottom-up attention computation}\label{subsec13}
Bottom-up attention refers to an attentional effect that is naturally evoked by the content of an image, thereby stimulating the human eye in a spontaneous manner. In order to accurately quantify this attention, we proposed a method involving the computation of multiple saliency maps. First, we adopted the frequency-tuned (FT) salient region detection method to output full resolution saliency maps with well-defined boundaries of salient objects \cite{achanta2009frequency}. The FT method can retain substantially more frequency content on the medical image. The calculation formula of the FT method is shown as:
\begin{equation}
	\mathbf{I}_{\mathbf{FT}}=S(x, y)=\left\|\mathbf{I}_{\mu}-\mathbf{I}_{\omega_{h c}}(x, y)\right\|
	\label{eq5}
\end{equation}
where $\mathbf{I}_{\mu}$ is the arithmetic mean image feature vector of the original image, and $\mathbf{I}_{\omega_{h c}}(x, y)$ is the corresponding image pixel value through the difference of Gaussian (DoG) filter \cite{man1982computational}. Here, the maximum cut-off frequency $\omega_{hc}$ was pre-set to $\pi$/2.75, which were demostrated to preserve the satisfactory high-fequency content from the original image \cite{achanta2009frequency}.

Physiologically,objects that are closer to the clinician's gaze point are more likely to be deemed important. Thus, we applied another fast minimum barrier distance (MBD) based saliency map calculation method \cite{zhang2015minimum} to simulate this physiological response characteristic. The MBD method can be directly applied to raw pixel values without any region abstraction, and acquires strong robustness of Pixel value fluctuation and excellent saliency extraction performance. However, one limitation is that regions near the image boundary can not be easily captured by the MBD method for saliency objects, which is in line with the visual gaze scene. The minimum MBD of any pixel $p_t$ can be is calculated through multiple iterations by following formula:
\begin{equation}
	\mathcal{D}\left ( p_t \right ) =\max \left\{\begin{matrix}\mathcal{D}\left ( p_t \right )
		\\ \max \{\mathcal{U}(p_a),\mathcal{I}(p_t) \} - \min \{\mathcal{L}(p_a),\mathcal{I}(p_t)\}
	\end{matrix}\right.
	\label{eq6}
\end{equation}
where $p_a$ represents the adjacent point of $p_t$, $\mathcal{I}(p_t)$ denotes the pixel value of $p_t$, and $\mathcal{U}(p_a)$ and $\mathcal{L}(p_a)$ are the highest and the lowest pixel values among the adjacent points of $p_a$ respectively. Noting that $\mathcal{D}(p_t)$ is the MBD of the point $p_t$ and is initialized to infinity before the first iteration. Then, the final iteration of $\mathbf{I}_{\mathbf{BMD}}$ is the saliency of the target image. Subsequently, we can obtain a combined saliency map calculation as:
\begin{equation}
	\mathbf{I}_{\mathbf{COM}}=\gamma \cdot \mathbf{I}_{\mathbf{FT}}+\eta \cdot \mathbf{I}_{\mathbf{BMD}}
	\label{eq7}
\end{equation}

\subsection{Segmentation on SAM with generated prompt points}\label{subsec14}
Recent study on the application of SAM in various image segmentation scenarios has showcased its superiority, particularly in terms of its zero-shot performance. The success of the SAM is largely based on the huge amount of annotated data. However, constructing an annotated dataset similar to SAM for a specific application poses a significant challenge. A practical approach is to utilize the the powerful zero-shot segmentation capability of SAM   and focus on optimizing the prompt points for an accurate detection of small lesions in fundus. For this end, we first construct an $N\times N$ grid like point coordinate set, represented by $\mathbf{P}_{\mathbf{grid}}$. $N$ is the number of points sampled along one side of the target image, which control the basic sampling resolution. Then, we extract the coordinates of salient points (denoted as $\mathbf{P}_{\mathbf{com}}$ ) from the binarized image. Eventually, we take the intersection of $\mathbf{P}_{\mathbf{grid}}$ and $\mathbf{P}_{\mathbf{com}}$ to obtain the final prompt points, which is computed as
\begin{equation}
	\mathbf{P}_{\mathbf{prompt}}= \mathbf{P}_{\mathbf{grid}} \cap  \mathbf{P}_{\mathbf{com}}
	\label{eq8}
\end{equation}

\subsection{Optimizing lesion segmentation through DKF}\label{subsec15}
The SAM with optimized prompt points can output multiple objects including suspected lesions and other false positives. Since the SAM did not conduct special learning about the knowledge of target lesions, we proposed a domain knowledge filtering (DKF) module to focus on the target lesions. The design criteria for the DKF module follows three aspects: shape, color and texture. According to the recognized medical knowledge, we digitized the description of the target lesion to construct the DKF module. 

\textbf{\emph{Shape}}. Inspired by \cite{sui2012four}, shape feature can be evaluated through roundness error which is described in \textcolor{blue}{(\ref{eq10})}. Clinically, micro-lesions, such as MAs, are in their early stages of growth, and are minimally affected by surrounding tissue structures. Hence, they typically exhibit nearly round shape and the lower bound of the roundness value was pre-set at 0.8 in this study.  
\begin{align}
    \mathrm{Round}(S,C) &= 4 \pi S / C^2 \nonumber \\
    &= \dfrac{4 \pi (\pi r^2)}{(2 \pi r)^2} && \text{when } S = \pi r^2 \text{ \& } C = 2 \pi r \nonumber \\
    &= 1
    \label{eq10}
\end{align}

\textbf{\emph{Color}}. By utilizing the LAB color space \cite{international2004image}, which closely aligns with the human eye's perception of colors, we can effectively express colors of target lesions and further reduce the occurrence of false positive lesions. For instance, microangiomas exhibit a higher red component compared to other lesions, while exudates consist of a larger percentage of yellow component than other types. Therefore, we only consider those with the red component accounting for the largest proportion as potential MAs.
    
\textbf{\emph{Texture}}. Clinically, the imaging manifestations of the identified micro lesions have a relatively narrow range of color values, resulting in a more delicate texture. Thus, we used the difference in pixel values to calculate the smoothness of local suspected lesions, which is indicated by	
\begin{equation}
    \text{Smoothness}=\sqrt{\dfrac{1}{N} \sum_{i=1}^{N}(I_i - \overset{-}{I})^2}
    \label{eq11}
\end{equation}	
where $I_i$ is the grayscale value in the target region and $\overset{-}{I}$ is the mean value. 
In general, the mico lesion region exhibits a smoother texture compared to the entire ROI or the attention region. We assume that micro lesions satisfy the condition of possessing a smoothness value lower than that of the surrounding background.
In summary, the final output comprises the micro lesions that satisfy the DKF criteria for shape, color, and texture simultaneously.

\begin{table}[b!]
	\caption{The quantity distribution of training and test set on the IDRiD and Retinal-Lesions datasets.}
	\begin{center}
		\label{tab: data}
		\vspace{-0.5em}
		\resizebox{0.95 \columnwidth}{10mm}{
			\begin{tabular}{cccccc} 
				\toprule                  
				\multirow{2}{*}{Dataset} & \multicolumn{2}{c}{IDRiD} && \multicolumn{2}{c}{Retinal-Lesions}  \\
				\cmidrule{2-3} \cmidrule{5-6}
				& Image-level & Lesion-level & & Image-level & Lesion-level \\
				\midrule 
				Train & 54 & 2403 & & 200 & 1197 \\
				Test & 27 & 1085 & & 135 & 750 \\
				\bottomrule 
			\end{tabular}
		}
	\end{center}
	\vspace{-2.0em}
\end{table}

\begin{figure}[t]
	\centering
	\includegraphics[width=0.95\columnwidth]{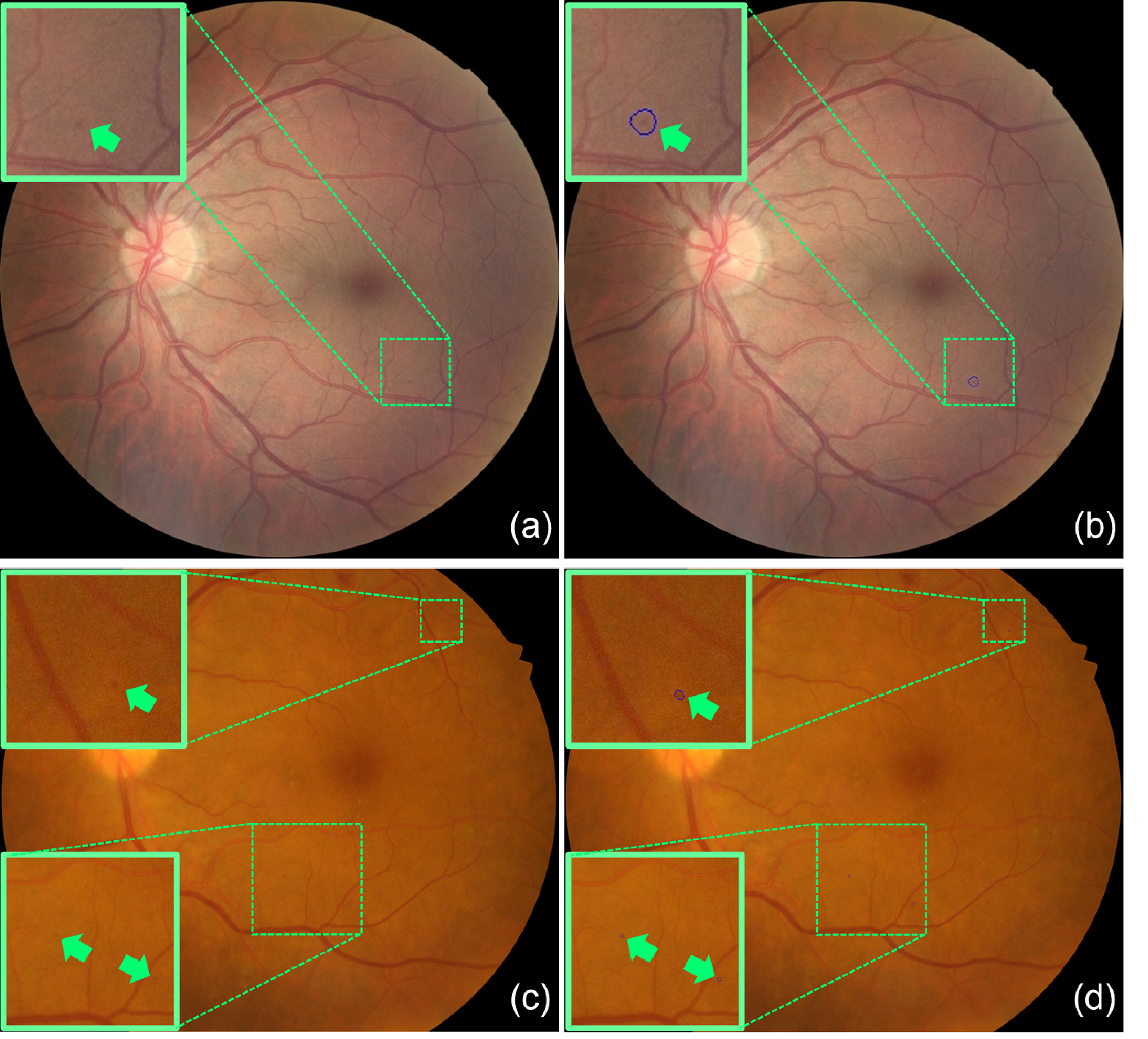}
	\vspace{0.0em}
	\caption{The first row comprises the original fundus image (a) and its corresponding annotated image (b), both originating from Retinal-Lesions dataset. Similarly, the second row is derived from IDRiD dataset. Inside, green solid box is the enlarged view of the inconspicuous small lesions on the fundus image, indicating the difficulty of the early fundus diseases detection tasks.}
	\label{fig:example_case}
	\vspace{-1.50em}
\end{figure}

\section{DATA AND IMPLEMENTATION DETAILS}
\subsection{Data description}
Two publicly available datasets, including IDRiD and Retinal-Lesions dataset, were adapted to validate the effectiveness of our proposed \textcolor{black} {GlanceSeg}. Specifically,
IDRiD dataset \cite{porwal2018indian} is acquired from a DR screening research. It consists of 81 pixel-level annotated color fundus images with signs of DR. The images have resolution of 4288$\times$2848 pixels and are stored in jpg file format. We randomly selected a set of 54 images, containing 100 microangioma lesions, for training stage, while the remaining 27 images, containing 50 microangioma lesions, were utilized for testing stage. 

Retinal-Lesions dataset \cite{wei2021learn} provides 315 fundus images, with microangioma lesions marked by either ellipses or polygons. Here, two images from the origin dataset were discarded due to damaged content within the images. In this database, 200 images were selected as the training set, while 100 images were for the test set. The input image is resized to 896$\times$896, as small lesions may not be discernible at a lower resolution. \textcolor{blue}{Fig. \ref{fig:example_case}} individually showcase examples of original images and lesion-annotated images from the two above databases. The data distribution of the two databases is illustrated in \textcolor{blue}{Table \ref{tab: data}}. 

\subsection{Implementation details}
Our study was implemented on a single NVIDIA RTX 3090 GPU and the official checkpoint of the ViT-B SAM model was utilized to conduct the segmentation of small microangioma lesions in both zero-shot and annotation-based fine-tuned settings. In the gaze maps-guided approximate lesion areas, we investigated the segmentation performance of the SAM-based foundation model by utilizing prompt points generated through diverse testing strategies. Regarding the comparative study methods, we conducted an extensive review of the performance of supervised learning approaches in published literature that utilized public IDRiD dataset. Additionally, we compared the zero-shot performance of SAM-based methods across various prompt points and assessed the effectiveness of fine-tuning SAM using the same training set of IDRiD dataset. We opted for the Area Under the Precision-Recall Curve (AUPR) and Dice coefficient as the evaluation metrics for our study.

\section{RESULTS ANALYSIS}

\begin{table*}[b]
    \footnotesize
	\caption{Comparison of accuracy (Acc), time consumption (Time), and attention dispersion score (ADS) in diabetic retinopathy (DR) diagnosis among three ophthalmologists.}\label{tab_DR5_without_GlanceSeg}
	\begin{tabular*}{\textwidth}{@{\extracolsep\fill}cccccccccc}
		\toprule%
		& \multicolumn{3}{@{}c@{}}{Ophthalmologist 1} & \multicolumn{3}{@{}c@{}}{Ophthalmologist 2} & \multicolumn{3}{@{}c@{}}{Ophthalmologist 3} \\\cmidrule{2-4}\cmidrule{5-7}\cmidrule{8-10}%
		Grading & Acc (\%) & Time (s) & ADS & Acc (\%) & Time (s) & ADS  & Acc (\%) & Time (s) & ADS\\
		\midrule
		DR 0  & 43.3 & 26.84 & 20.92 & 41.9 & 40.86 & 16.06 & 63.3 & 22.45 & 23.89\\
		DR 1  & 50.0 & 23.04 & 16.28 & 60.0 & 31.33 & 15.02 & 83.3 & 21.42 & 22.53\\
		DR 2  & 73.3 & 22.80 & 15.99 & 83.3 & 21.64 & 20.90 & 70.0 & 18.73 & 18.34\\
		DR 3  & 73.3 & 19.94 & 22.33 & 66.6 & 14.56 & 21.29 & 76.6 & 16.42 & 24.41\\
		DR 4  & 83.3 & 22.77 & 18.93 & 86.6 & 13.39 & 14.82 & 86.6 & 11.88 & 19.06\\
		\bottomrule 
	\end{tabular*}
\end{table*}

\subsection{Analysis of clinicians's reviewing \& diagnosing image}\label{subsec2}
To leverage the advantages of \textcolor{black}{GlanceSeg}, we carefully analyzed the characteristics of ophthalmologists during their diagnosing DR in the fundus image. We invited three ophthalmologists of two and five-year clinical experience to participate in our research. Ophthalmologists were provided a same set of fundus images to perform the task of grading DR into five scales, including health (DR0), mild (DR1), moderate (DR2), severe (DR3), and proliferative (DR4). Based on 30 randomly selected images for each grade in the Retinal-Lesions dataset, we compared the average grading accuracy, the time consumption, and the attention dispersion score among three ophthalmologists. As shown in \textcolor{blue}{Table \ref{tab_DR5_without_GlanceSeg}}, the ophthalmologist 3 with five-year experience generally demonstrates a higher diagnostic accuracy and required less time than the other two junior ophthalmologists of two-year experience. More importantly, for the two junior ophthalmologist, their reviewing on DR0 and DR1 took a longer time with relatively lower diagnostic accuracy than their performance on other DR grades. We speculated that more severe DR levels possess prominent pathological features which are easily recognizable and beneficial to the diagnostic accuracy. Therefore, this work concentrates on the more challenging task of assisting in the diagnosis of DR1 solely based on the presence of small microangiomas.    

\begin{figure*}[t]
	\centering
	\includegraphics[width=1.8\columnwidth]{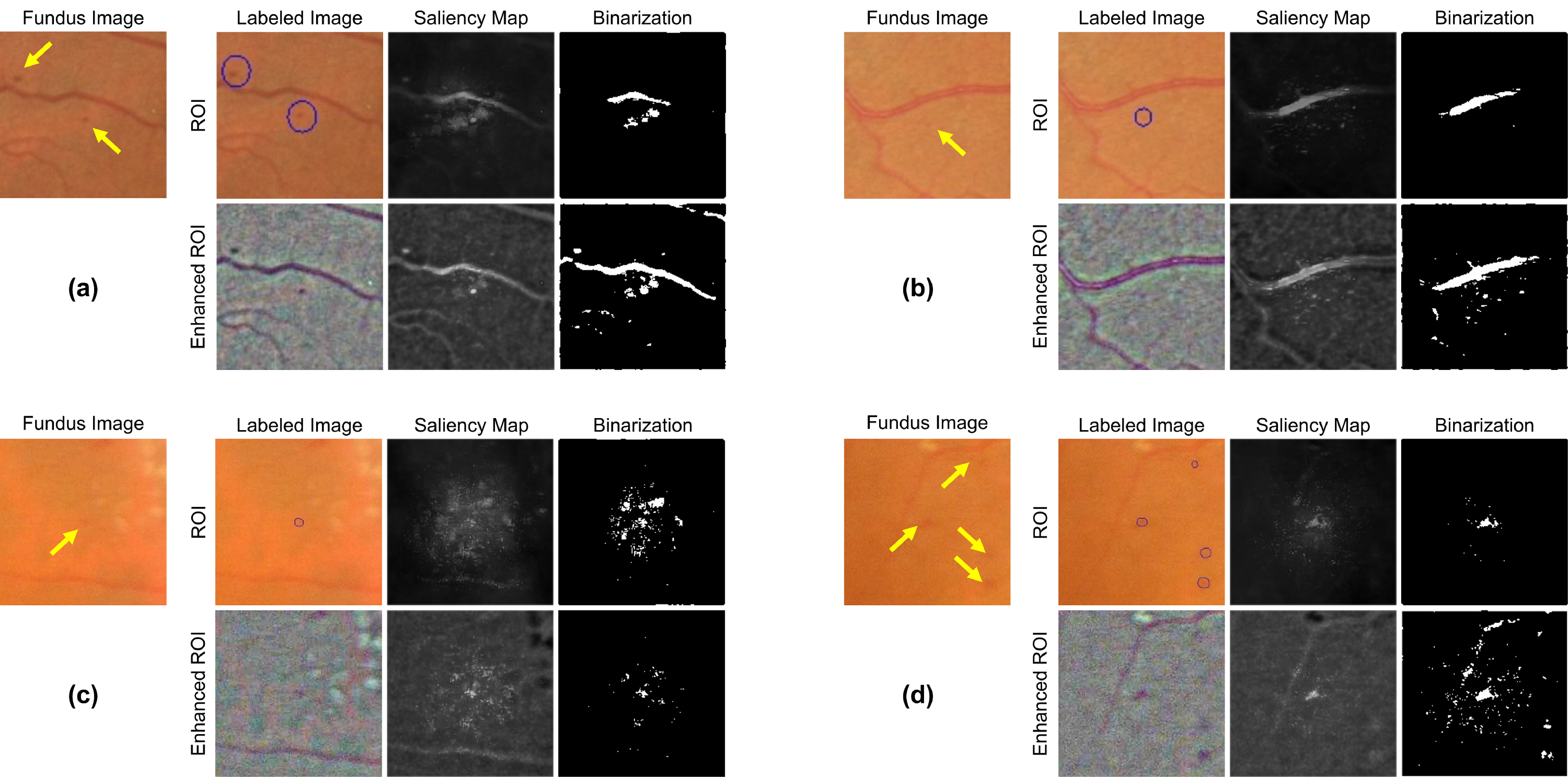}
	\vspace{0.0em}
	\caption{(a) and (b) are illustrative examples from Retinal-Lesions while (c) and (d) are from  IDRiD dataset. The first row of each example represents the extracted fundus regions of interest (ROI) guided by eye-tracking, the saliency map of the ROI, and the post-processed map through binarization, sequentially. The second row is analogous to the first row, with the distinction that the ROI has undergone preliminary image enhancement processing for optimization. Yellow arrows highlight the microangioma lesions, which serve as early indicators for the diagnosis of diabetic retinopathy.}
	\label{fig:enh-sm}
	\vspace{-1.0em}
\end{figure*}

\begin{figure*}[t!]
	\centering
	\includegraphics[width=1.6\columnwidth]{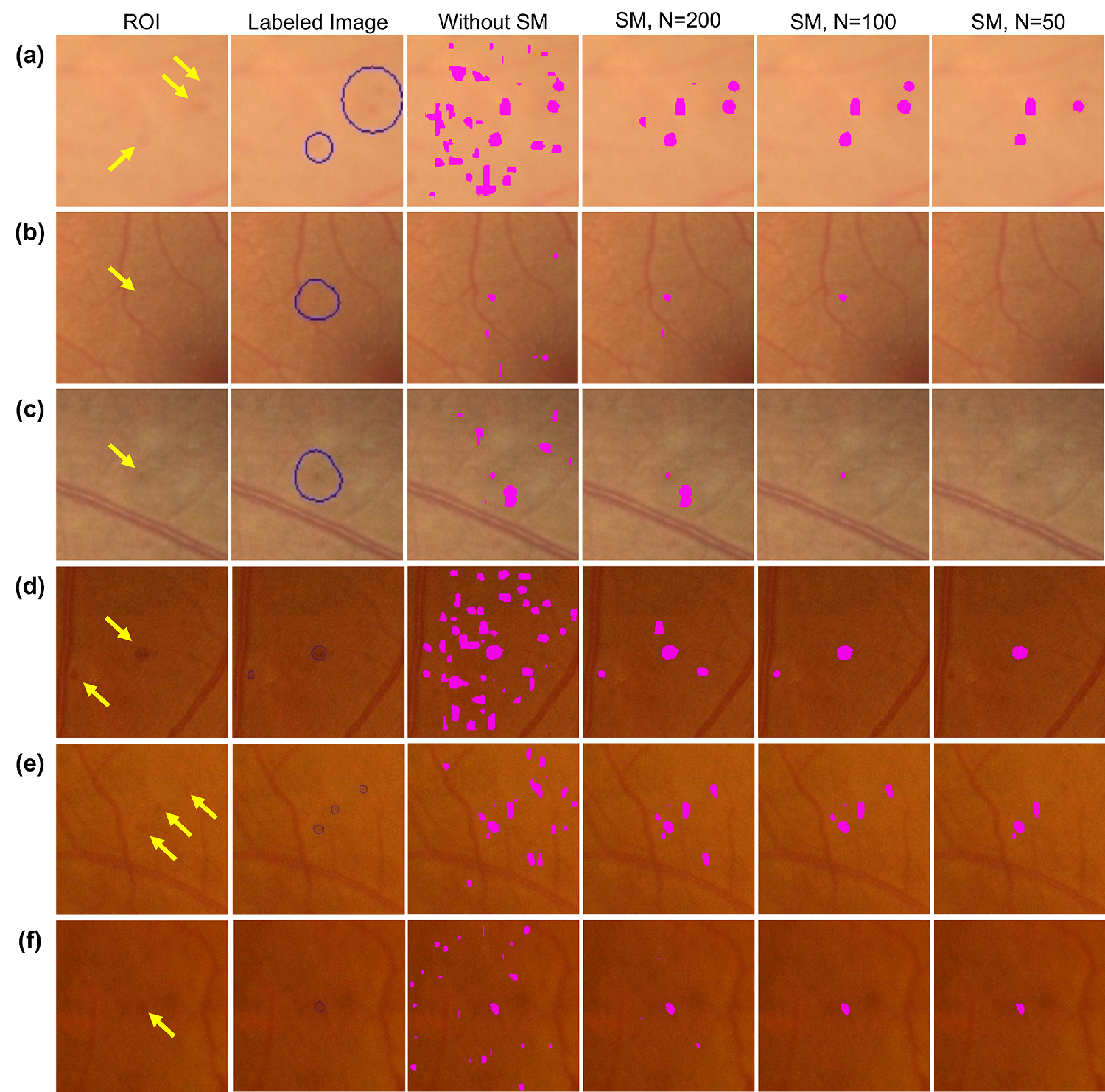}
	\vspace{0.0em}
	\caption{Row-wise instance display: showcasing intermediate segmentation of microangioma lesions with or without saliency maps under diverse hyperparameter configurations. Inside, SM is an abbreviation for saliency map.}
	\label{fig:sm-nvalue}
	\vspace{0.0em}
\end{figure*}

\begin{figure*}[t!]
	\centering
	\includegraphics[width=1.75\columnwidth]{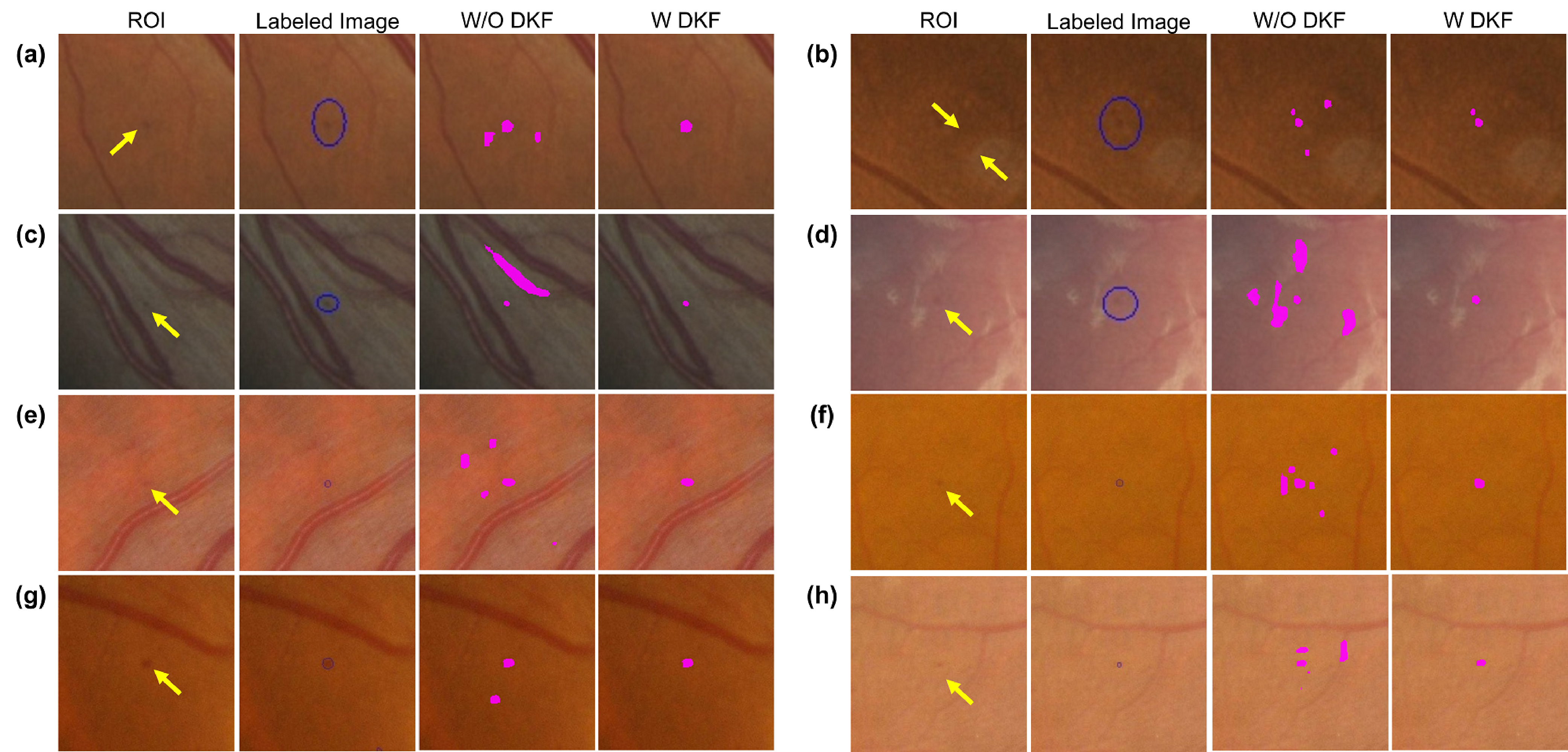}
	\vspace{0.0em}
	\caption{Enhanced purification of microangioma lesion segmentation results through domain knowledge filter post-processing technique. DKF represents the domain knowledge filter.}
	\label{fig:dkf}
	\vspace{0.0em}
\end{figure*}

\subsection{Visualizing the zero-shot performance of GlanceSeg}\label{subsec3}
Based on the eye-tracker, the clinician's attention (i.e., gaze points) helps us focus on the ROI where the micro lesions are located. In clinical practice, besides some salient lesions clinicians would generally observe the locations where lesions are prone to occur according to their medical knowledge and experience. After calculating the amount of attention, we cropped a group of rectangular ROIs acquiring high attention from each image. Then, each ROI was enhanced for achieving a better saliency map. The effect of the calculated saliency map is shown in \textcolor{blue}{Fig. \ref{fig:enh-sm}}.
We assumed that tiny lesions represent the significant foregrounds, and extracted saliency maps which further ensures to generate prompt points through sampling. Specifically, the saliency maps were produced through the image fusion.

We uniformly sampled $N\times N$ points along the $X$ and $Y$ axes from the saliency map and the points sampled to the salient area are used as the prompt points of the SAM. Afterwards, we obtained the segmentation results of multiple salient object, including suspected lesions. We evaluated the segmentation outcomes at varying sample sizes, namely \(N = 50\), 100, and 200, to compare their performance.
The examples of the segmentation results of the SAM are displayed in \textcolor{blue}{Fig. \ref{fig:sm-nvalue}}. We can see that a \textcolor{black}{larger} value of $N=200$ is more prone to yielding false negatives, whereas a \textcolor{black}{smaller} value of $N=50$ is inclined to generate false positives. An appropriate $N=100$ is crucial for achieving improved segmentation lesions.

So far, we have not introduced any domain knowledge and only utilized the zero-shot ability of the SAM. The medical prior knowledge can help us further improve the zero-shot performance of the small lesion detection. Thus, using the proposed domain knowledge filtering, some false positives can be ruled out obviously which is shown in \textcolor{blue}{Fig. \ref{fig:dkf}}.

\begin{table}
        \caption{Performance summary of the existing supervised methods on public A dataset, demonstrating the superiority of our proposed GlanceSeg both with or without supervision.}
        \vspace{-0.5em}
	\begin{center}
		\label{tab_comparison}%
		\begin{tabular}{@{}lcccccc@{}}
			\toprule
			Method & Learning type & AUPR & DICE\\
			\midrule
			U-Net\cite{ronneberger2015u}    & Supervised & 0.4504 & 0.1624 \\        
			iFLYTEK\cite{porwal2020idrid}    & Supervised & 0.5017  & - \\
			RTN\cite{huang2022rtnet}    & Supervised & 0.4897   & - \\
			MCA-UNet\cite{wang2023mca}    & Supervised & 0.5206  & 0.3850 \\
			CLC-Net\cite{wang2023clc}    & Supervised & 0.5460 & - \\
			SAM\cite{kirillov2023segment}     & zero-shot & 0.1762  & 0.0553 \\
			GlanceSeg (ours) & zero-shot & 0.5523  & 0.3680\\
			GlanceSeg+ (ours) & fine-tune & \textbf{0.5705} & \textbf{0.3944}\\
			\bottomrule 
		\end{tabular}
	\end{center}
	\vspace{-0.50em}
\end{table}

\subsection{Comparison with state-of-the-art methods}\label{subsec4}
To demonstrate the superiority of our proposed GlanceSeg framework, we have reviewed some  segmentation performance of supervised learning methods based on the public IDRiD dataset. These supervised learning method were trained on the train set and tested on the test set of IDRiD dataset. From \textcolor{blue}{Table \ref{tab_comparison}}, due to the limited training set, which consists of 54 fundus images covering a total of 2,403 annotated microangiomas, the AUPR values of the supervised learning methods are approximately 0.5. The traditional UNet has the poorest performance, with an AUPR of 0.4504, while the CLC-Net network exhibits the best performance, achieving an AUPR of 0.5460. The average performance of supervised learning methods also demonstrates that the tiny microangioma segmentation is a challenging task. \textcolor{blue}{Table \ref{tab_comparison}} also displays the performance of SAM and our proposed GlanceSeg, both of which are unsupervised learning methods. 
SAM performs poorly in terms of segmentation under prompt points based on $200\times200$ uniformly sampled points from the original image, with an AUPR of only 0.1762. This suggests that SAM is not readily suitable for tiny microangioma segmentation in the medical applications. Surprisingly, our proposed GlanceSeg method, without requiring supervised information, has achieved a notably high AUPR of 0.5523, showcasing its effectiveness in the micro-lesion segmentation scenario. Furthermore, we utilized the training set to fine-tune our proposed GlanceSeg, achieving a significant performance improvement with AUPR of \textcolor{black}{0.5705} and DICE score of \textcolor{black}{0.3944}. 

\begin{figure}[t]
	\centering
	\includegraphics[width=1\columnwidth]{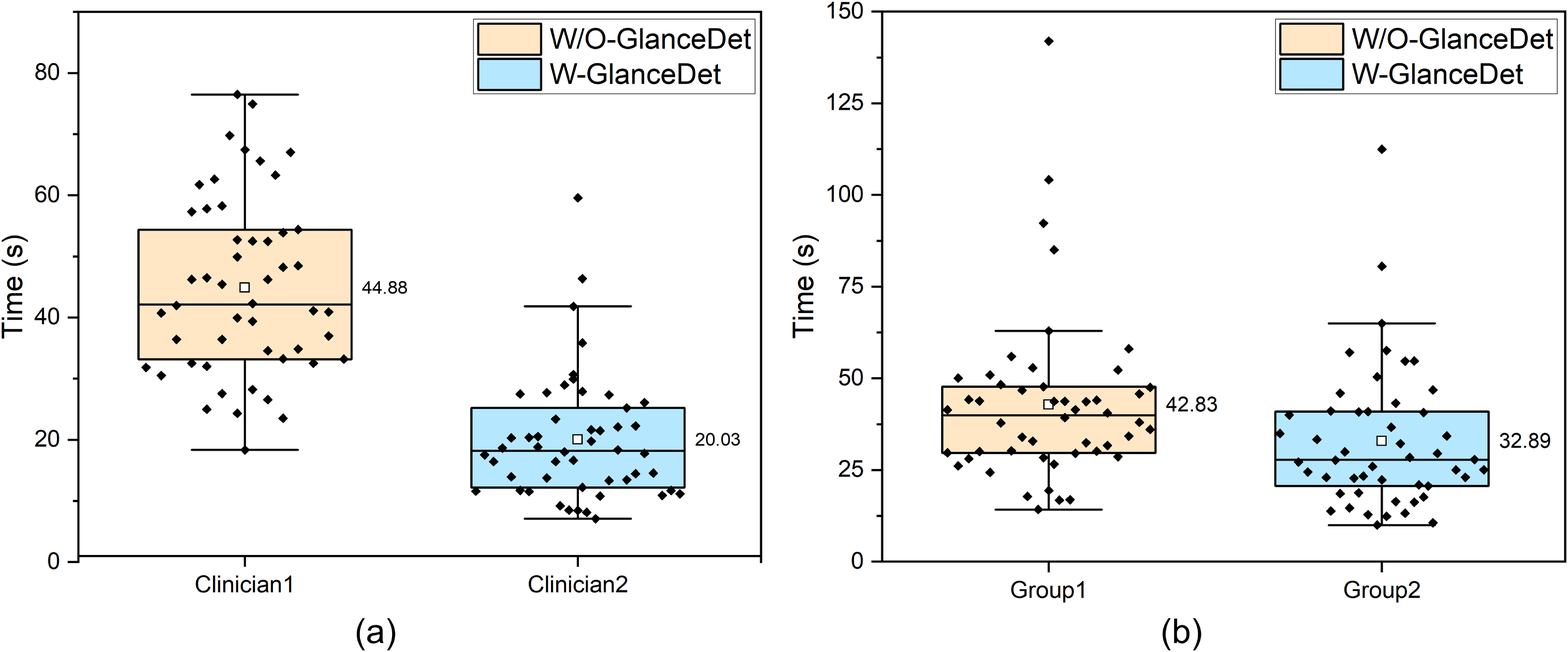}
	\vspace{-1.5em}
	\caption{(a) Comparing the annotation time of two clinicians with comparable years of experience, clinician B with GlanceSeg assistance vs. clinician A without assistance. (b) Comparing a clinician's annotation time on two closely distributed data groups of equal size. Group 2 employed GlanceSeg-assisted annotation, while Group 1 did not benefit from this assistance.}
	\label{fig:boxPlot}
	\vspace{-1em}
\end{figure}

\subsection{Evaluating the annotation efficiency of GlanceSeg}\label{subsec5}
During the evaluation of retinal fundus images by ophthalmologists, GlanceSeg is capable of presenting the suspected segmented microangioma lesions in real-time, following the guidance of their eye gaze maps.
Therefore, GlanceSeg is beneficial to pixel-level annotation of microangioma lesions on fundus images, potentially reducing the time burden on clinicians. 
Accordingly, we designed two sets of experiments to demonstrate the role of GlanceSeg in micro-lesions annotation and the average time taken by the two junior ophthalmologists for annotating each image is depicted in \textcolor{blue}{Fig. \ref{fig:boxPlot}(a)}. Two junior ophthalmologists with comparable expertise were tasked with annotating microangioma lesions in a set of 50 fundus images with the DR1 label from the Retinal-Lesions dataset. We can observe that, when aided by GlanceSeg, clinician B's average time consumption is 20.03 seconds, which is less than half of the time taken by clinician A when not utilizing GlanceSeg. 

In addition, one senior ophthalmologist with five years of expertise was requested to annotate two sets of 50 DR1-labeled fundus images, which were randomly selected from the Retinal-Lesions dataset. From \textcolor{blue}{Fig. \ref{fig:boxPlot}(b)}, it can be observed that the average time taken by the Group 2 utilizing GlanceSeg is 32.89 seconds, while the control Group 1 requires 42.83 seconds. Meanwhile, when compared with \textcolor{blue}{Fig. \ref{fig:boxPlot}(b)}, the senior has a more advantageous average time consumption in the same annotation setting.   
Consequently, we can conclude that GlanceSeg effectively enhances the annotation efficiency for pixel-level tiny lesions for both senior and junior clinicians.

\begin{figure*}[t!]
	\centering
	\includegraphics[width=1.5\columnwidth]{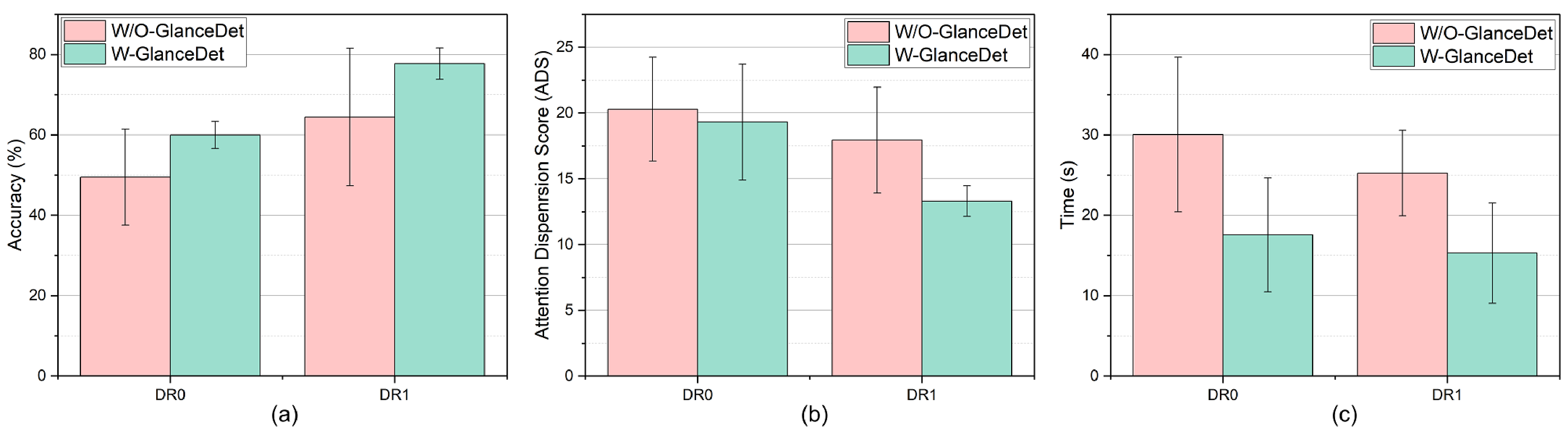}
	\vspace{-0.8em}
	\caption{Comparative average performance metrics for DR0 and DR1 diagnosis: Evaluating the clinician w/o GlanceSeg assistance.}
	\label{fig:three_indicators}
	\vspace{-0.5em}
\end{figure*}

\begin{table*}[b]
    \footnotesize
	\caption{Evaluating GlanceSeg-Assisted DR Diagnosis: A Statistical Analysis of Accuracy (ACC), Time Consumption (Time), and Attention Dispersion Score (ADS).}\label{tab_comparison_GlanceSeg_DR5}
	\begin{tabular*}{\textwidth}{@{\extracolsep\fill}cccccccccc}
		\toprule%
		& \multicolumn{3}{@{}c@{}}{Ophthalmologist 1} & \multicolumn{3}{@{}c@{}}{Ophthalmologist 2} & \multicolumn{3}{@{}c@{}}{Ophthalmologist 3} \\\cmidrule{2-4}\cmidrule{5-7}\cmidrule{8-10}%
		Grading & Acc (\%) \(\uparrow\) & Time (s) \(\downarrow\) & ADS \(\downarrow\) & Acc (\%) \(\uparrow\) & Time (s) \(\downarrow\) & ADS \(\downarrow\)  & Acc (\%) \(\uparrow\) & Time (s) \(\downarrow\) & ADS \(\downarrow\)\\
		\midrule
		\multirow{2}{*}{DR 0}  & 56.6 & 14.32 & 18.36 & 63.3 & 25.69 & 15.44 & 60.0 & 12.67 & 24.10\\
                  & {\scriptsize\(\textcolor{blue}{\uparrow\ 13.30}\)}
                  & {\scriptsize\(\textcolor{red}{\downarrow\ 12.52}\)}   
                  & {\scriptsize\(\textcolor{red}{\downarrow\ 2.56}\)} 
                  & {\scriptsize\(\textcolor{blue}{\uparrow\ 21.40}\)}  
                  & {\scriptsize\(\textcolor{red}{\downarrow\ 15.17}\)}
                  & {\scriptsize\(\textcolor{red}{\downarrow\ 0.62}\)}  
                  & {\scriptsize\(\textcolor{red}{\downarrow\ 3.30}\)}  
                  & {\scriptsize\(\textcolor{red}{\downarrow\ 9.78}\)}  
                  & {\scriptsize\(\textcolor{blue}{\uparrow\ 0.21}\)} \\
		\multirow{2}{*}{DR 1}  & 73.3 & 12.61 & 13.32 & 80.0 & 22.44 & 12.13 & 80.0 & 10.84 & 14.47\\
                  & {\scriptsize\(\textcolor{blue}{\uparrow\ 23.30}\)}
                  & {\scriptsize\(\textcolor{red}{\downarrow\ 10.43}\)}   
                  & {\scriptsize\(\textcolor{red}{\downarrow\ 2.96}\)} 
                  & {\scriptsize\(\textcolor{blue}{\uparrow\ 20.00}\)}  
                  & {\scriptsize\(\textcolor{red}{\downarrow\ 8.89}\)}
                  & {\scriptsize\(\textcolor{red}{\downarrow\ 2.89}\)}  
                  & {\scriptsize\(\textcolor{red}{\downarrow\ 3.30}\)}  
                  & {\scriptsize\(\textcolor{red}{\downarrow\ 10.58}\)}  
                  & {\scriptsize\(\textcolor{red}{\downarrow\ 8.06}\)} \\ 
		\multirow{2}{*}{DR 2}  & 66.6 & 14.89 & 16.97 & 80.0 & 20.59 & 18.63 & 70.0 & 11.83 & 18.34\\
                  & {\scriptsize\(\textcolor{red}{\downarrow\ 6.70}\)}
                  & {\scriptsize\(\textcolor{red}{\downarrow\ 7.91}\)}   
                  & {\scriptsize\(\textcolor{blue}{\uparrow\ 0.98}\)} 
                  & {\scriptsize\(\textcolor{red}{\downarrow\ 3.30}\)}  
                  & {\scriptsize\(\textcolor{red}{\downarrow\ 1.05}\)}
                  & {\scriptsize\(\textcolor{red}{\downarrow\ 2.27}\)}  
                  & {\scriptsize\(\textcolor{black}{--}\)}  
                  & {\scriptsize\(\textcolor{red}{\downarrow\ 6.90}\)}  
                  & {\scriptsize\(\textcolor{black}{--}\)} \\
		\multirow{2}{*}{DR 3}  & 76.6 & 18.37 & 18.15 & 70.0 & 16.08 & 18.39 & 80.0 & 12.74 & 19.45\\
                  & {\scriptsize\(\textcolor{blue}{\uparrow\ 3.30}\)}
                  & {\scriptsize\(\textcolor{red}{\downarrow\ 1.57}\)}   
                  & {\scriptsize\(\textcolor{red}{\downarrow\ 4.18}\)} 
                  & {\scriptsize\(\textcolor{blue}{\uparrow\ 3.40}\)}  
                  & {\scriptsize\(\textcolor{blue}{\uparrow\ 1.52}\)}
                  & {\scriptsize\(\textcolor{red}{\downarrow\ 2.90}\)}  
                  & {\scriptsize\(\textcolor{blue}{\uparrow\ 3.40}\)}  
                  & {\scriptsize\(\textcolor{red}{\downarrow\ 3.68}\)}  
                  & {\scriptsize\(\textcolor{red}{\downarrow\ 4.96}\)} \\
		\multirow{2}{*}{DR 4}  & 83.3 & 14.23 & 15.83 & 83.3 & 15.04 & 15.69 & 90.0 & 10.73 & 13.94\\
                  & {\scriptsize\(\textcolor{black}{--}\)}
                  & {\scriptsize\(\textcolor{red}{\downarrow\ 8.54}\)}   
                  & {\scriptsize\(\textcolor{red}{\downarrow\ 3.10}\)} 
                  & {\scriptsize\(\textcolor{red}{\downarrow\ 3.30}\)}  
                  & {\scriptsize\(\textcolor{blue}{\uparrow\ 1.65}\)}
                  & {\scriptsize\(\textcolor{blue}{\uparrow\ 0.87}\)}  
                  & {\scriptsize\(\textcolor{blue}{\uparrow\ 3.40}\)}  
                  & {\scriptsize\(\textcolor{red}{\downarrow\ 1.15}\)}  
                  & {\scriptsize\(\textcolor{red}{\downarrow\ 5.12}\)} \\
		\bottomrule 
	\end{tabular*}
\end{table*}

\begin{figure}[t!]
	\centering
	\includegraphics[width=1\columnwidth]{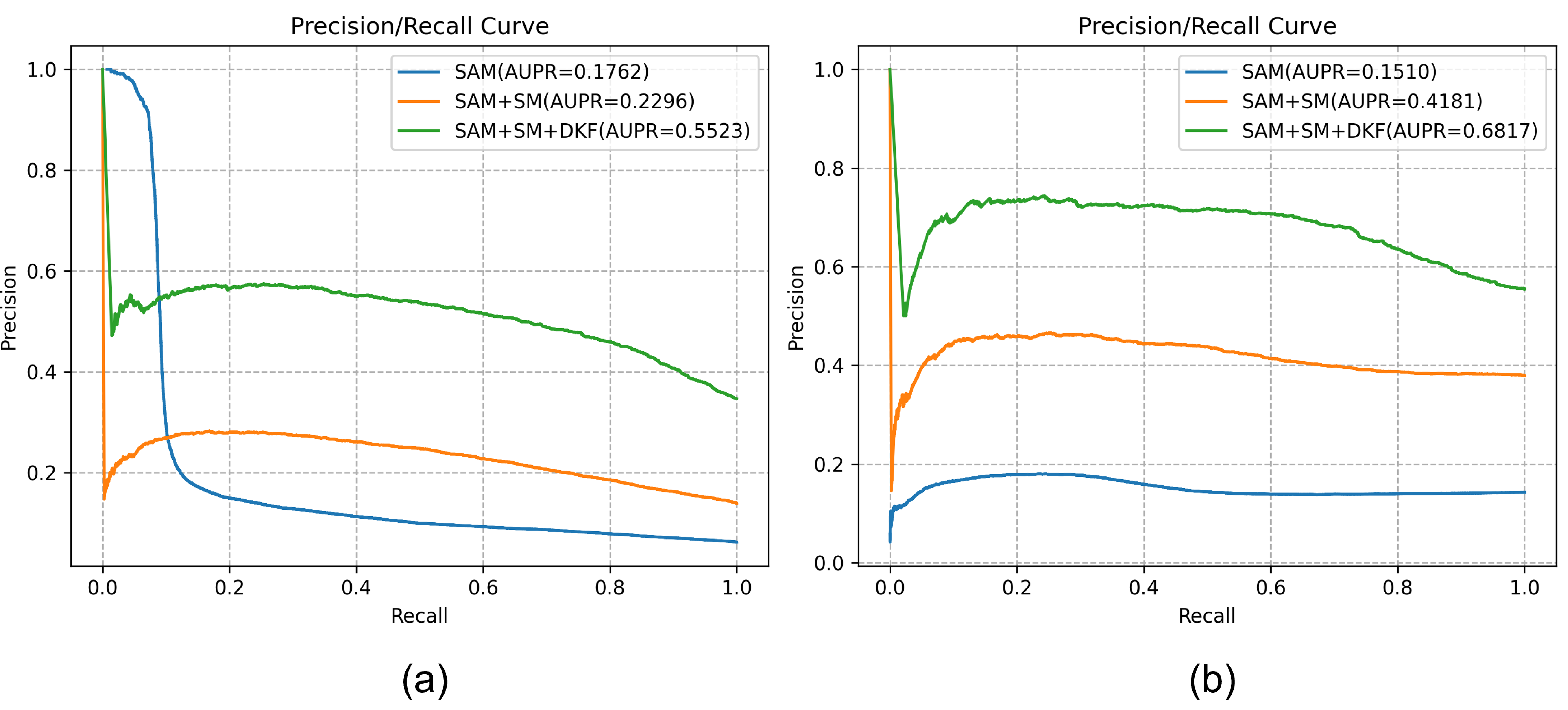}
	\vspace{-1.4em}
	\caption{Comparison of Precision/Recall curve plots among three ablation experiment groups on IDRiD (a) and Retinal-Lesions (b) datasets. Inside, SM is an abbreviation for saliency map while DKF represents the domain knowledge filter. }
	\label{fig:AUPR}
	\vspace{-1em}
\end{figure}

\begin{figure*}[t!]
	\centering
	\includegraphics[width=1.5\columnwidth]{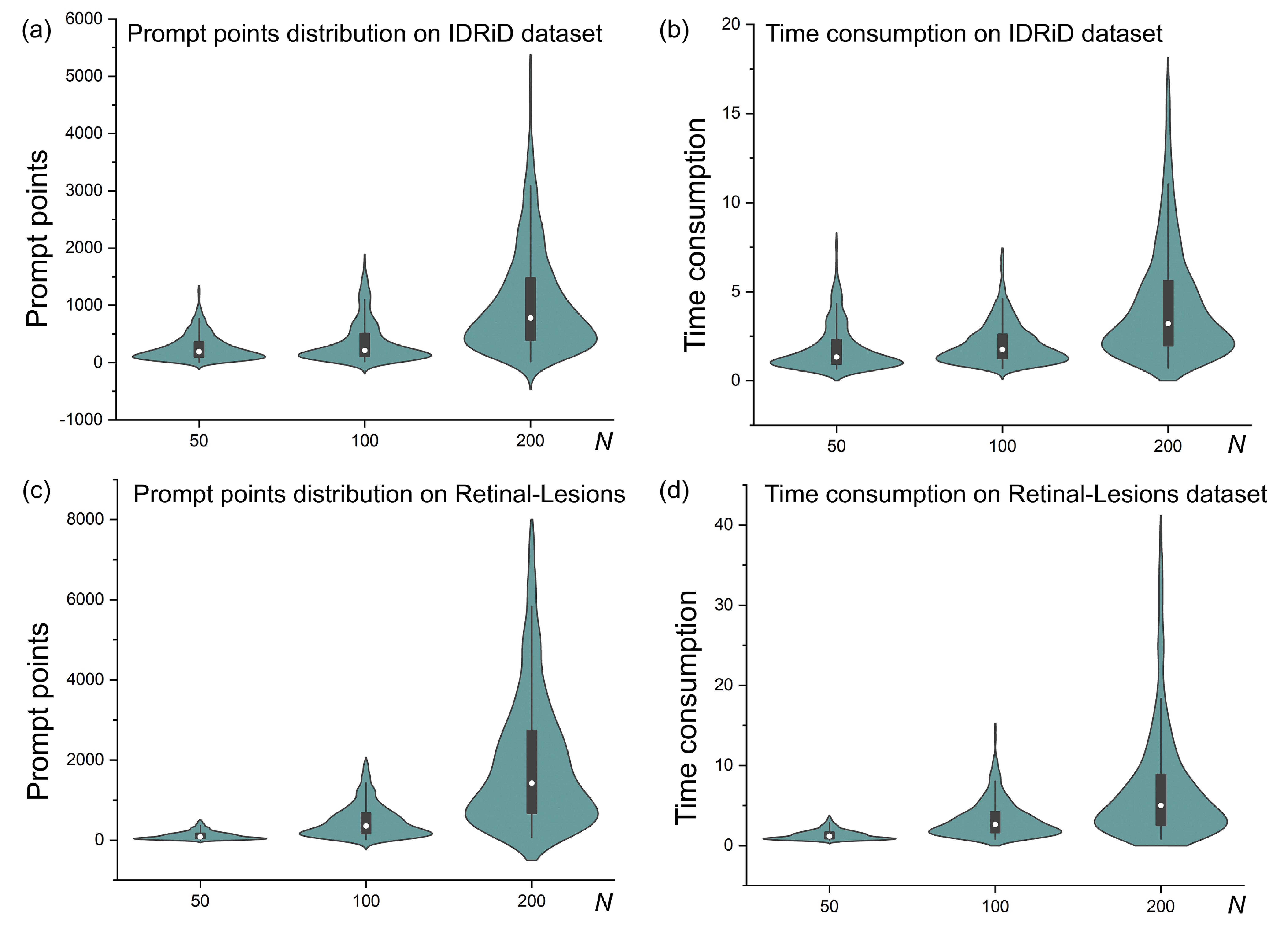}
	\vspace{-0.5em}
	\caption{Violin plot comparing prompt points and model inference time for two datasets under different sample point dimension sizes (N).}
	\label{fig:POINTS-TIME-Number}
	\vspace{-0em}
\end{figure*}

\begin{figure}[t!]
	\centering
	\includegraphics[width=1.0\columnwidth]{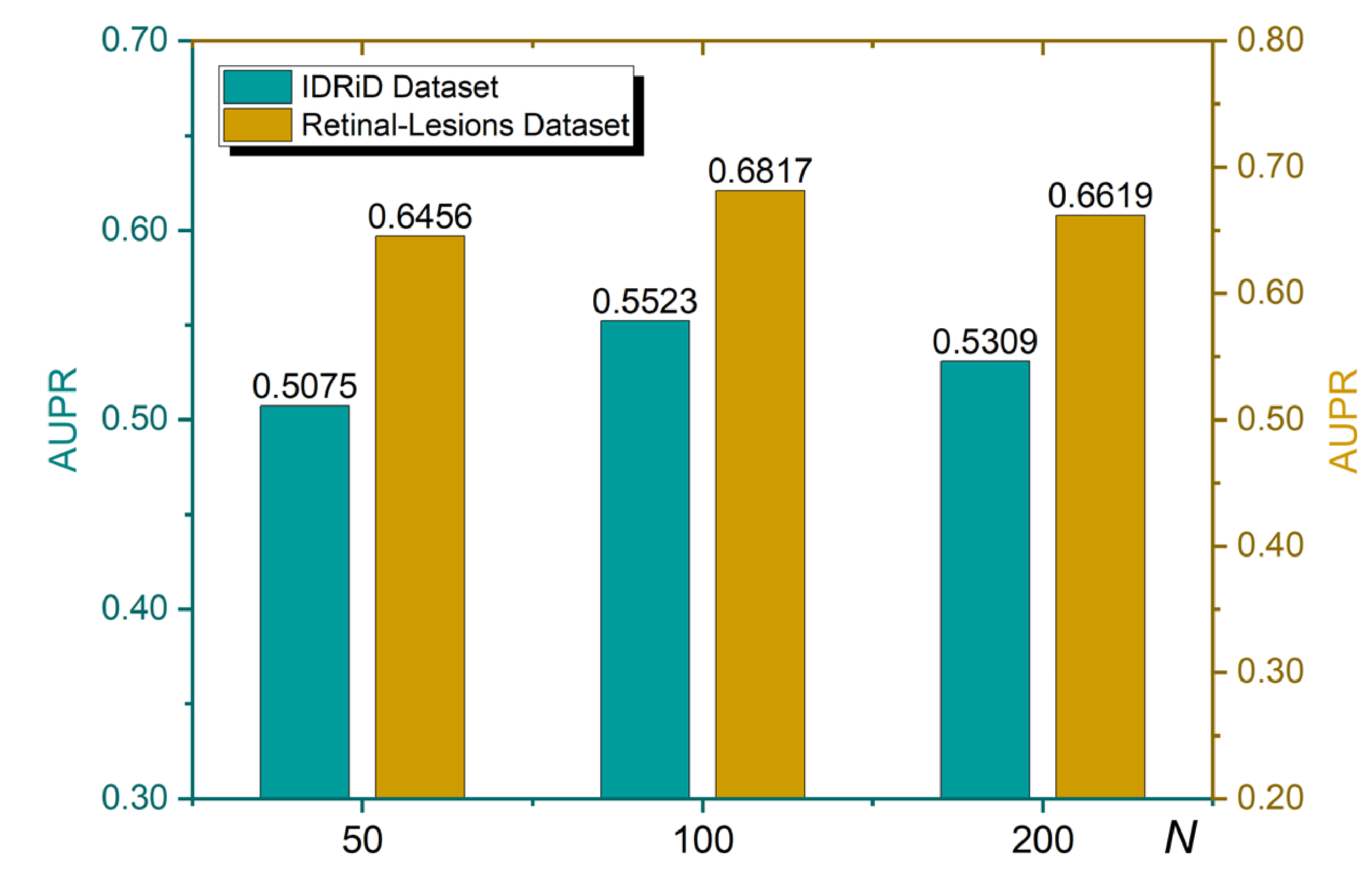}
	\vspace{-1.5em}
	\caption{Comparison analysis of AUPR based on two above-mentioned datasets under different hyperparameters: investigating the impact of Sample prompt Point dimension size (N).}
	\label{fig:AUPR-Number}
	\vspace{-1.5em}
\end{figure}

\subsection{Evaluating the diagnostic performance of GlanceSeg}\label{subsec6}
In this section, we conducted a performance analysis of the GlanceSeg-assisted diagnosis approach. Two junior and one senior ophthalmologists, aided by GlanceSeg, assessed 30 images from each category within Retinal-Lesions dataset. The mean accuracy, time consumption, and attention disperse score for every level are presented in \textcolor{blue}{Table \ref{tab_comparison_GlanceSeg_DR5}}. Note that, there is no overlap between the test data in this section and the data in \textcolor{blue}{Table \ref{tab_DR5_without_GlanceSeg}}; however, both data sets are randomly selected without overlap from the same dataset and demonstrate comparable distributions. From \textcolor{blue}{Table \ref{tab_comparison_GlanceSeg_DR5}}, for junior ophthalmologists (1 \& 2), GlanceSeg is capable of enhancing the accuracy of early DR grading (DR0 \& DR1), reducing the time required for diagnosis, and concurrently lowering the attention disperse score. It indicates that GlanceSeg can assist in locating the small lesions, enabling the junior ophthalmologists to precisely pinpoint the impaired area, thereby improving diagnostic efficiency. For the senior ophthalmologist, during the early diagnosis stage of DR, GlanceSeg also shortens the diagnostic time; however, the diagnostic accuracy does not increase, but rather exhibits a minor decrease. \textcolor{black}{We speculated that senior clinicians, possessing extensive experience in reviewing images, demonstrate an enhanced sensitivity to even the minute lesions. Consequently, GlanceSeg improves the efficiency of image review, yet its contribution to performance is not significant. Given that there is no overlap between the two sets of test data, it is reasonable to observe the slight fluctuations in performance.}  
As GlanceSeg aims primarily segmenting minor lesions, it does not offer significant benefits for the non-early DR classification, such as DR2, DR3, and DR4 cases, where both the number and area of lesions are pronounce.

\section{Discussion}
\subsection{Ablation study} 
GlanceSeg, grounded on the foundation model SAM, enables unsupervised real-time segmentation of retinal microangioma lesions in fundus images. GlanceSeg primarily consists of three core modules: gaze map-guided coarse segmentation of peri-microangiomal regions, microangioma segmentation via saliency map-generated prompt points based on the SAM model, and further  segmentation refinement utilizing domain knowledge filter. Gaza maps of ophthalmologists during fundus image interpretation contribute to roughly localize the region of tiny lesions, narrowing the segmentation scope and ensuring real-time segmentation. Due to the remarkably small size of microangioma lesions, it is unfeasible to accomplish lesion segmentation under unsupervised conditions using the SAM model without the benefit of gaze maps. 

Then, ablation studies were conducted to demonstrate the effectiveness of prompt points derived from the saliency map and optimization of segmentation outcomes using the domain knowledge filter. \textcolor{blue}{Fig. \ref{fig:AUPR}(a)} and \textcolor{blue}{(b)} individually showcase the Precision/Recall curves for three ablation experiments conducted on IDRiD and Retinal-Lesions datasets. From \textcolor{blue}{Fig. \ref{fig:AUPR}(a)}, the SAM group refers to the results obtained following an initial segmentation process based on the gaze map, utilizing 200$\times$200 evenly distributed sampling points as prompt points for further segmentation. The introduction of prompt points derived from the saliency map has improved the SAM model's performance, with the AUPR increasing from 0.1762 to 0.2296.  
Furthermore, the domain knowledge filter group enhances performance by effectively eliminating segmentation outcomes that diverge from well-founded medical prior knowledge, achieving an AUPR of 0.5523.
\textcolor{blue}{Fig. \ref{fig:AUPR}(b)} draws a similar conclusion regarding the Retinal-Lesions dataset, with the increases from the domain knowledge filter being more prominent. 
In summary, the ablation studies demonstrate that both components (i.e., prompt points generated by the saliency map and the domain knowledge filter) contribute to the performance improvements, and the combination of them can achieve the optimal performance. 

\subsection{Hyperparameter sensitivity analysis}
During the generation of our tailored prompt points, $N \times N$ points are uniformly sampled from the saliency map. In this section, we investigated the impact of the hyperparameter $N$ on the quantity of prompt points, the SAM model's inference time consumption, and the effectiveness of microangioma segmentation. \textcolor{blue}{Fig. \ref{fig:POINTS-TIME-Number}} depicts the violin plots comparing the quantity of prompt points and the model inference time for two datasets utilized in the study, under varying sample point dimension sizes ($N$). 
The violin plot, a data visualization technique, integrates features from both box plots and kernel density plots, thereby offering a holistic depiction of prompt points and inference time consumption distribution under a range of $N$ settings. From \textcolor{blue}{Fig. \ref{fig:POINTS-TIME-Number}}, as $N$ increases, there is a corresponding growth in the number of prompt points, which in turn leads to extended inference time. It is rational that the intersection points (i.e., prompt points) between $N \times N$ sampling points and the saliency map may increase as $N$ grows, leading to an extended processing time. \textcolor{blue}{Fig. \ref{fig:AUPR-Number}} displays the AUPR across two datasets under various $N$ settings. As $N$ decreases, the number of prompt points is reduced, potentially causing the omission of microangioma segmentation and resulting in false negatives. Conversely, with an increase in $N$, the quantity of prompt points rises, which may lead to over-segmentation and the introduction of noise, ultimately producing false positives. Hence, for both utilized datasets, the optimal AUPR is attained when $N$ equals 100, indicating a moderate size. In IDRiD dataset, the peak AUPR is 0.5523, whereas in Retinal-Lesions dataset, it reaches 0.6817. We empirically set $N$ as 100 in the experiments. 

\section{Conclusion}
This paper presents a real-time, label-free diagnosis approach for the early detection of DR using a foundation model SAM. The proposed approach incorporates gaze maps obtained during clinicians' image evaluation, enabling top-down attention-driven coarse lesion localization. It is further enhanced by bottom-up attention guidance through prompt points derived from saliency maps. Additionally, a domain knowledge filter is applied to enhance the segmented microangioma of the foundation model. The effectiveness of the proposed approach, named GlanceSeg, is validated through qualitative and quantitative results obtained from two newly-reconstructed public datasets. The results demonstrate that GlanceSeg significantly improves both diagnostic performance and annotation efficiency for early DR.
Furthermore, the study highlights that GlanceSeg can further enhance segmentation performance through fine-tuning with annotations. This finding suggests that incorporating GlanceSeg-assisted annotations for model fine-tuning holds promise as a research direction to achieve continual learning capabilities and exceptional performance on early DR detection.

\bibliographystyle{IEEEtran}
\bibliography{refs}

\end{document}